\documentclass[journal=mamobx, manuscript=article]{achemso}

\usepackage{eurosym}
\usepackage{enumitem}
\usepackage{tabularx}
\usepackage{multirow}
\usepackage{graphics}
\usepackage{epsfig}
\usepackage{multirow}
\usepackage{amssymb}
\usepackage{amsmath}
\usepackage{leftidx}
\usepackage{epsfig}
\usepackage{color,soul}
\usepackage{array}
\usepackage{array, makecell}
\usepackage[explicit]{titlesec}
\usepackage{comment}
\usepackage{xcolor}

\title{Scaling Law for Sequence-Induced Demixing \\
of Compositionally Identical Copolymers}

\author{Artem M. Rumyantsev}
\email{rumyantsev@ncsu.edu}
\author{Alexey A. Gavrilov}%
\affiliation{Department of Chemical and Biomolecular Engineering, North Carolina State University, Raleigh, North Carolina 27695-7905, USA}
\date{\today}

\begin{document}

\begin{tocentry}
\includegraphics[scale=0.17]{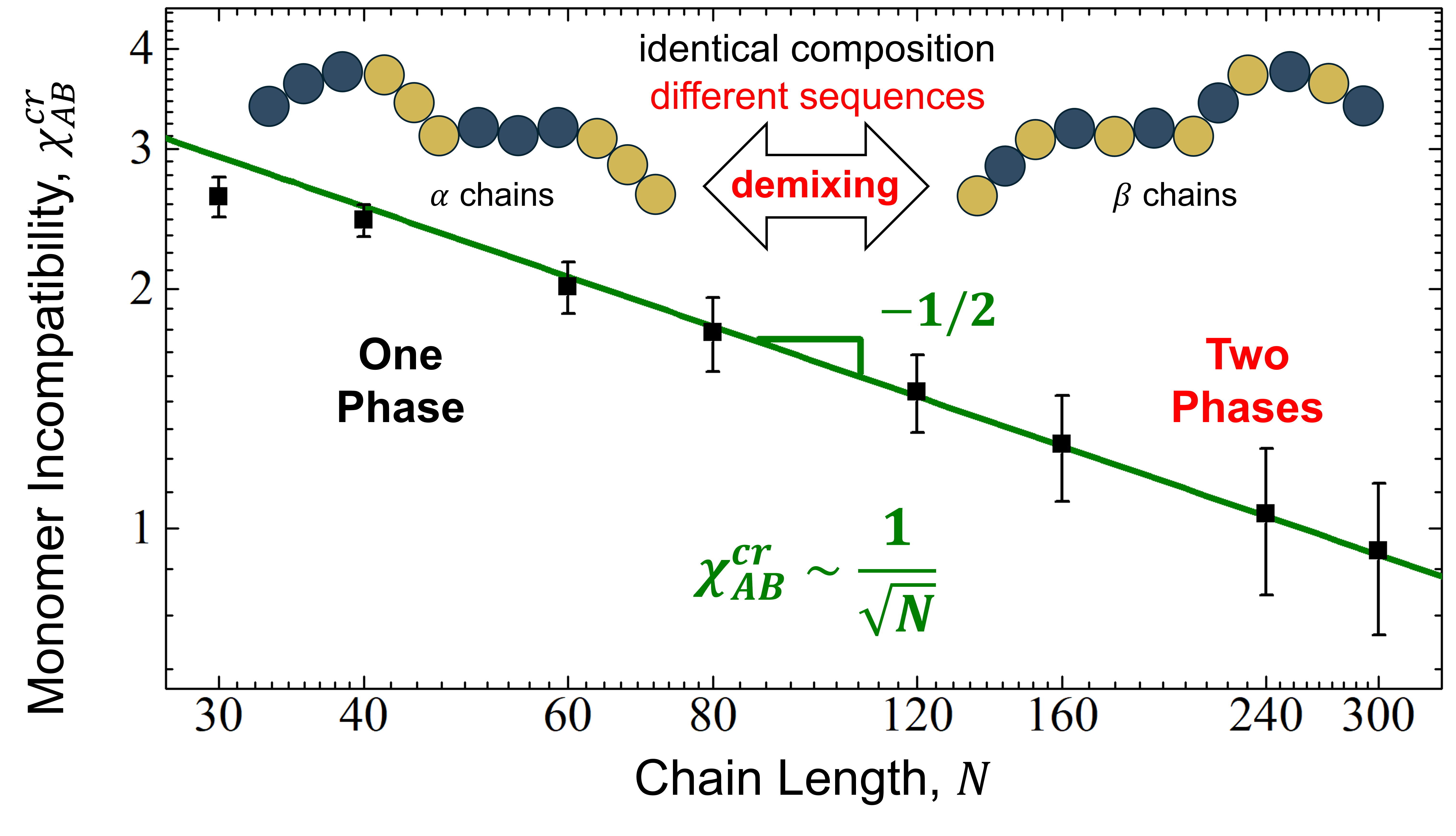}
\end{tocentry}

\begin{abstract}

The critical incompatibility of polymers with different compositions scales inversely with their length. For instance, a mixture of A and B homopolymers of length $N$ segregates at $\chi_{AB}^{cr} = 2/N$. But what if the difference between the blend components is subtler? We demonstrate that a mixture of AB copolymers with identical composition --- equal amounts of A and B monomers --- but different primary sequences can still phase separate. Incompatibility arises from distinct positional correlations between monomers of different chains. Calculating the Gaussian fluctuation correction to the free energy reveals that critical incompatibility from sequence differences follows a distinct yet universal scaling with chain length, $\chi_{AB}^{cr} \sim 1 / \sqrt{N}$. This power law holds for both regular-sequence and statistical copolymers. A closed-form expression is derived for blends of block-alternating chains. The new theoretical scaling is confirmed by coarse-grained simulations, offering important insights into multiphase coexistence in biomolecular condensates.

\end{abstract}

\newpage

Macroscopic phase separation in polymer blends is a classical problem in polymer physics. The minimal theoretical model considers the mixture of A and B homopolymers of equal length $N$, with the Flory-Huggins (FH) interaction parameter $\chi_{AB}$ serving as the measure of monomer immiscibility. Since the seminal works by Flory and Huggins~\cite{huggins-1942, flory-1942, flory-book}, it is well established that the critical point coordinate scales inversely with chain length~\cite{gennes-book, GK-book, RC-book}:  
\begin{equation}
    \chi_{AB}^{cr} = \frac{2}{N}
\label{eq:start}
\end{equation}
The slope of $-1$ represents a cornerstone result: it is highly universal and holds for copolymer mixtures whenever they differ in chemical composition~\cite{scott-1952, ten-brinke-1983}.

\begin{figure}[t]
\includegraphics[width = 5.0in]{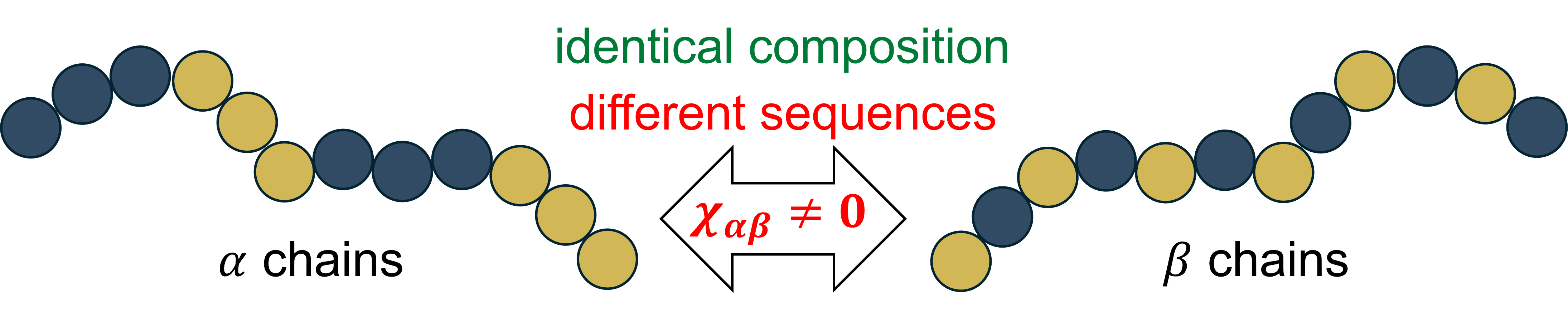}
\caption{
\label{fig:1} 
Compositionally identical AB copolymers with different primary sequences, $\alpha$ and $\beta$. Effective incompatibility between them, $\chi_{\alpha\beta} \neq 0$, arises due to correlation effects.}
\end{figure}

But what if the blend components have identical overall composition and \textit{only} differ in the primary sequence of their monomers, as shown in Figure~\ref{fig:1}? Classical theories, such as the FH lattice model, fail to address this question and erroneously predict ideal miscibility in such cases.

In this Letter, we show that copolymers with identical composition can phase-separate \textit{solely} due to differences in their primary sequences. We develop a general theory of this effect, which predicts a distinct slope of $-1/2$ in the dependence of critical monomer incompatibility on the copolymers' length: 
\begin{equation}
    \chi_{AB}^{cr} \sim \frac{1}{\sqrt{N}}
\end{equation}
This universal and elegant scaling law, valid across different sequence types, is further confirmed by coarse-grained simulations.

\textbf{Sequence-Induced Incompatibility.} We consider two compositionally identical AB copolymers, denoted $\alpha$ and $\beta$, each comprising $50\%$ of A and B monomers, but with different primary sequences, as shown in Figure 1. Both copolymers are flexible, with the monomer (segment) size $a$. In a solvent-free blend, the volume fractions of copolymers add to unity, $\phi_{tot} = \phi_{\alpha} + \phi_{\beta} = 1$. 

Mean-field (MF) approaches, such as the FH lattice theory, correspond to the saddle point approximation of polymer field theory \cite{F-book}. They completely neglect positional correlations between A and B monomers due to their connectivity, which are different between the monomers of $\alpha$ and $\beta$ chains due to their different primary sequences. Instead, the interaction between the chains is approximated by the sum of interactions between their disjointed monomers~\cite{gennes-book, GK-book, RC-book}. Due to the identical chemical composition, the resulting MF FH interaction parameter between monomers of $\alpha$ and $\beta$ chains is exactly zero, $\chi_{\alpha \beta}^{MF} = 0$. 

Thus, sequence-induced immiscibility is a subtle effect that cannot be described within the saddle point approximation. In simple terms, long blocks prefer to be surrounded by other long rather than short blocks --- and $\alpha$ chains favor other $\alpha$ chains over $\beta$ chains --- to maximize the energy gain from positional correlations, i.e., the efficiency of A/B avoidance. The longer the blocks, the more effective this avoidance, as connectivity enforces cooperativity in their mutual repulsion: long blocks in a highly blocky $\alpha$ copolymer repel each other more strongly than isolated A/B monomers in nearly alternating $\beta$ copolymer. 

The underlying physics of this effect is, in spirit, analogous to charge correlations in a Debye–Hückel plasma (an electrolyte solution). In a naive MF picture, the average electrostatic energy would vanish, since each charge is surrounded by an equal number of positive and negative ions. In reality, however, the plasma develops non-trivial positional correlations: opposite charges are more likely to be found near one another, whereas like charges are statistically kept apart --- an effect known as Debye screening. These correlations generate a finite, negative correlation energy \cite{Landau-book, Levin2002}.

In the present work, we go beyond MF and calculate the Gaussian fluctuation correction to the free energy using the well-established framework of random phase approximation (RPA) \cite{gennes-book}. This enables us to find the effective incompatibility between $\alpha$ and $\beta$ chains and predict regularities of sequence-induced macroscopic phase separation.

Within the RPA, the difference in the primary sequence manifests in the different single-chain structure factors $g (q)$ for AA, AB, and BB pairs of monomers between $\alpha$ and $\beta$ chains \cite{leibler-1980}. For homopolymers, the well-known quantity $g (q)$ is the Fourier representation of the single-chain correlation function; it captures the spatial correlations of monomers along a polymer, which directly set the scattering intensity observed from an isolated chain. For $\alpha$ chains, the structure factors are denoted $g_{AA}^{\alpha} (q)$, $g_{AB}^{\alpha} (q)$, and $g_{BB}^{\alpha} (q)$.  Owing to the compositional symmetry of both copolymers, $g_{AA}^{\alpha} (q) = g_{BB}^{\alpha} (q)$ and $g_{AA}^{\beta} (q) = g_{BB}^{\beta} (q)$. The matrix of bare structure factors 
\begin{equation}
\textbf{G}_{0} (q) = 
   \begin{pmatrix}
   \phi_{\alpha} g_\text{AA}^{\alpha} (q) + \phi_{\beta} g_\text{AA}^{\beta} (q)  &
   \phi_{\alpha} g_\text{AB}^{\alpha} (q) + \phi_{\beta} g_\text{AB}^{\beta} (q) \\
   \phi_{\alpha} g_\text{AB}^{\alpha} (q) + \phi_{\beta} g_\text{AB}^{\beta} (q) &
   \phi_{\alpha} g_\text{AA}^{\alpha} (q) + \phi_{\beta} g_\text{AA}^{\beta} (q) \\
   \end{pmatrix}
\label{struct-matrix}
\end{equation}
reflects the connectivity-induced intramolecular AA, AB, and BB positional correlations in the blend, weighted by composition \cite{leibler-1980, erukh-1982}. Here and below, boldface symbols correspond to $2\times2$ matrices.

The total structure factor of all the monomers in the blend
is equal to the sum of all the elements of $\textbf{G}_{0} (q)$ matrix:
\begin{equation}
    S(q) = \sum_{i = \alpha, \beta} \phi_{i} \Big[ g_{AA}^{i} (q) + 2 g_{AB}^{i} (q) + g_{BB}^{i} (q) \Big] =
    \frac{12}{ \left( q a \right)^2}
\label{eq:S}
\end{equation}
Here we assumed that both copolymers are sufficiently long, $N_{i} \gg 1$ for $i = \alpha, \beta$, and approximated their structure factors given by the Debye functions~\cite{gennes-book, GK-book, RC-book} with that of an infinitely long Gaussian coil.

We also define the matrix determinant, $W (q) = \det \textbf{G}_{0} (q) = S(q) R(q) / 2$, in the spirit of the theory of block-copolymers~\cite{leibler-1980, erukh-1982}. The introduced function
\begin{equation}
    R(q) = \phi_{\alpha} \Big[ g_{AA}^{\alpha} (q) - g_{AB}^{\alpha} (q) \Big] + \phi_{\beta} \Big[ g_{AA}^{\beta} (q) - g_{AB}^{\beta} (q) \Big]
\label{eq:R-def}
\end{equation}
depends on the blend composition and the primary sequence of the copolymers.

The key idea of the RPA is that the positional correlations between the monomers are defined by both their chemical connectivity and non-covalent interactions. Expressing this mathematically, the inverse matrix of the Fourier transforms of the monomer-monomer correlation functions in the copolymer blend 
\begin{equation}
    \textbf{G}^{-1} (q) = \textbf{G}_{0}^{-1} (q) + \textbf{U}(q)
\end{equation}
is equal to the sum of the structural part due to monomer connectivity, $\textbf{G}_{0}^{-1} (q)$, and the interaction part, $\textbf{U}(q)$. The first, structural contribution is obtained by inverting $\textbf{G}_{0} (q)$ matrix given by eq.~\ref{struct-matrix}, see Supporting Information for details. The interaction contribution is equal to the matrix of the Fourier transforms of the pairwise interaction potential between the monomers:
\begin{equation}
    \textbf{U}(q) = 
\chi_{AB}
\begin{pmatrix}
0 & 1 \\
1 & 0
\end{pmatrix}
+ \mathcal{B} 
\begin{pmatrix}
1 & 1 \\
1 & 1
\end{pmatrix}
\label{eq:U}
\end{equation}
It takes into account incompatibility between A and B monomers defined by the FH parameter $\chi_{AB}$ and blend incompressibility enforced by the second term with the infinite bulk modulus, $\mathcal{B} \to \infty$~\cite{morse-2010}.

We are now in a position to calculate the Gaussian fluctuation correction to the free energy using the RPA framework~\cite{F-book, edwards-1966, olvera-1988, erukh-1988, BF-1994, BLF-1994, joanny-2002, wang-2002, RJTdP-2022}. It accounts for fluctuations in the density fields of A and B monomers, $\phi_A(\vec{r})$ and $\phi_B(\vec{r})$, by integrating across the entire spectrum of wavevectors $q$ \cite{potemkin-2007}. The resulting correction expressed in $k_B T / a^3$ units equals 
\begin{equation}
    F_{RPA} = \frac{1}{2} \int \frac{d^3 q} {\left( 2 \pi / a \right)^3} 
    \ln \left[ \frac{\det \textbf{G}^{-1} (q)} 
    { \left. \det \textbf{G}^{-1} (q) \right|_{\chi_{AB} = 0} } \right] \\ 
    = \frac{1}{2} \int \frac{d^3 q} { \left( 2 \pi / a \right)^3 } 
    \ln \left[ 1 - \chi_{AB} R(q) \right] 
\label{eq:RPA-gen}
\end{equation}
where we made use of $\det \textbf{G}^{-1} (q) = 2 \mathcal{B} \left[ 1 - \chi_{AB} R(q) \right] / W (q) $. The result of eq.~\ref{eq:RPA-gen} corresponds to fluctuations in the local composition of the blend and is described by the order parameter $\psi(\vec{r}) = \phi_{A} (\vec{r}) - \phi_{B} (\vec{r})$ widely used in the theory of block-copolymers~\cite{leibler-1980, OK-1986, FL-1989}. Indeed, integral~\ref{eq:RPA-gen} diverges at the spinodal of microphase separation transition (MST) between A and B monomers given by
\begin{equation}
    \chi_{AB} R(q) = 2 \chi_{AB} \frac{W(q)}{S(q)} = 1
\end{equation}
Note that the introduced incompressibility constraint, $\mathcal{B} \to \infty$, completely suppresses fluctuations in the total density of A and B monomers providing its uniform value in blend, $\phi_{tot} (\vec{r}) = \phi_{A} (\vec{r}) + \phi_{B} (\vec{r}) = 1$, and the respective contribution vanishes. In semidilute solutions, this contribution would survive and appear additively as the Edwards correction independent of the copolymer sequences~\cite{rumyantsev-2024}.

For homogeneous blends far from the MST, $ \chi_{AB} R(q) \ll 1 $, compositional fluctuations are weak, providing the RPA applicability; that is, the Gaussian correction dominates the higher-order terms in the perturbative loop expansion~\cite{rumyantsev-2024}. This also permits expanding the logarithm in eq.~\ref{eq:RPA-gen}, yielding a series representation of the RPA correction in powers of the incompatibility $\chi_{AB}$: 
\begin{equation}
    F_{RPA} \approx  - \chi_{AB} \frac{a^3}{2} \int \frac{d^3 q} { \left( 2 \pi \right)^3 } R(q) 
    - \chi_{AB}^2 \frac{a^3}{4} \int \frac{d^3 q} { \left( 2 \pi \right)^3 } R^2(q)
\label{eq:F_RPA-series}
\end{equation}
Here, only the first two terms have been retained. In view of eq.~\ref{eq:R-def}, the first term is linear in the volume fractions of $\alpha$ and $\beta$ copolymers, $\phi_{\alpha}$ and $\phi_{\beta} = 1 -\phi_{\alpha}$. It provides zero contribution to the osmotic pressure and a constant one to the chemical potential, thereby not contributing to the blend thermodynamics, and should be omitted~\cite{Landau-book}. Because of the formal mathematical divergence of the linear term, this omission can be considered the field theory renormalization~\cite{rumyantsev-2024} analogous to that performed in refs.~\citenum{wang-2002}, \citenum{morse-2007}, and \citenum{rauscher-2023}. Physically, the divergent term can be interpreted as the self-energy of the compositional field $\psi (r)$.~\cite{rumyantsev-2024}

In the second term in eq.~\ref{eq:F_RPA-series}, contributions independent of and linear in the volume fraction $\phi_{\alpha}$ can also be omitted. As a result, one arrives at the fluctuation correction to the free energy quadratic in the volume fraction:
\begin{equation}
     F_{RPA}^{\psi} = \chi_{\alpha \beta}^{RPA} \phi_{\alpha} \phi_{\beta} =
    \chi_{\alpha \beta}^{RPA} \phi_{\alpha} \left(1 - \phi_{\alpha} \right)
\end{equation}
It suggests that the difference in the primary sequence of $\alpha$ and $\beta$ chains effectively generates \textit{pairwise repulsions} between their monomers, with the effective monomer-monomer FH incompatibility equal to
\begin{equation}
    \chi_{\alpha \beta}^{RPA} = \frac{a^3} {8 \pi^2} \chi_{AB}^2 
    \int_{0}^{\infty} \left( \frac{ \partial R(q) } { \partial \phi_{\alpha} } \right)^2 q^2 dq 
    \label{eq:chi-final}
\end{equation}
Here, the function
\begin{equation}
    \frac{ \partial R(q) } { \partial \phi_{\alpha} } = 
    g_{AA}^{\alpha} (q) - g_{AA}^{\beta} (q) -
    g_{AB}^{\alpha} (q) + g_{AB}^{\beta} (q) 
   \label{eq:R(q)}
\end{equation}
characterizes the sequence difference between $\alpha$ and $\beta$ copolymers. 
We emphasize that the result of eq.~\ref{eq:chi-final} is universal and holds for any periodic or random sequence with the characteristic block length much lower than the total chain length --- this requirement provides the independence of the structure factors, $\partial R(q) / \partial \phi_{\alpha}$ function, and monomer incompatibility $\chi_{\alpha\beta}^{RPA}$ on $N$. The physical reason for sequence-induced incompatibility is the different positional correlations experienced by the monomers belonging to $\alpha$ and $\beta$ chains, as seen from the structure of eq.~\ref{eq:R(q)}. 

In simple terms, positional correlations mean that A and B monomers avoid one another, thereby lowering the system’s free energy. This is consistent with the general principle of \textit{“like dissolves like”}. The subtlety is that this effect is correlation/fluctuation-induced and is therefore of second order, $\chi_{\alpha\beta}^{RPA} \sim \chi_{AB}^2 $, in contrast to the first-order incompatibility in A/B homopolymer blends with $\chi_{\alpha\beta} = \chi_{AB}$, where A monomers prefer to be surrounded by other A monomers already at the MF level.

\textbf{Block-Alternating Copolymers.}
To illustrate our general result and provide physical intuition for the role of sequence mismatch, consider the blend of block-alternating copolymers $\left( \text{A}_{m} \text{B}_{m} \right)_{N_{\alpha}/2m}$ and $\left( \text{A}_{n} \text{B}_{n} \right)_{N_{\beta}/2n}$ with the block lengths equal to $m$ and $n$ in $\alpha$ and $\beta$ chains, respectively. The single-chain structure factors for the former are given by~\cite{bates-2004, rumyantsev-2024}
\begin{equation}
    g_\text{AA}^{\alpha} (x_m) = \frac{m}{x_m^2} 
    \left[ x_m - \frac{1 - e^{-x_m}} {1 + e^{-x_m}} \right] 
\label{eq:str-fac-m}
\end{equation}
\begin{equation}
    g_\text{AB}^{\alpha} (x_m) = \frac{m}{x_m^2} 
    \left[ \frac{1 - e^{-x_m}} {1 + e^{-x_m}} \right]
\label{eq:str-fac-n}
\end{equation}
with $x_{m} = m \left( q a \right)^2 / 6$, and analogously with $m \to n$ substitution for the latter. They are independent of the chain lengths provided that the chains comprise numerous blocks, $N_{\alpha} \gg m$ and $N_{\beta} \gg n$.
Substituting them into eq.~\ref{eq:R(q)} would not enable analytical integration in eq.~\ref{eq:chi-final}. To find the effective incompatibility $\chi_{\alpha\beta}^{RPA}$ in the closed form, the fractions in eqs.~\ref{eq:str-fac-m} and \ref{eq:str-fac-n} for $n,m \gg 1$ can be approximated as
$(1-e^{-x}) / (1 + e^{-x}) \approx x / (2 + x)$ to obtain
\begin{equation}
\frac {\partial R(q)} {\partial \phi_{\alpha}} \approx 
\frac{m - n} { 2 \left( 1 + q^2 r_m^2 \right) \left( 1 + q^2 r_n^2 \right) }
\label{eq:dR/dF-approx}
\end{equation}
The introduced characteristic lengths $r_{m} = a \sqrt{m/12} $ and $r_{n} = a \sqrt{n/12}$ are proportional (equal up to the $\sqrt{2}$ factor) to the gyration radii of the blocks. Performing integration in eq.~\ref{eq:chi-final} yields the effective incompatibility between the monomers of $\alpha$ and $\beta$ chains that increases quadratically with the difference between the block lengths $m$ and $n$:
\begin{equation}
    \chi_{\alpha\beta}^{RPA} 
    \approx \frac{3 \sqrt{3}} {16 \pi} \frac{ \left( m - n \right)^2 } { \left( \sqrt{m} + \sqrt{n} \right)^{3} } \chi_{AB}^2
\label{eq:chi_eff_final}
\end{equation}
Thus, the greater \textit{the relative variation} in primary sequences, the stronger the copolymer immiscibility. Note that the relative error introduced to $\chi_{\alpha\beta}^{RPA}$ due to the approximation of eq.~\ref{eq:dR/dF-approx} does not exceed 40\%, as demonstrated by comparing to the exact result in eq. S7 of the Supporting Information.

\textbf{Scaling Law.}
We now return to the general case of $\alpha$ and $\beta$ copolymers with arbitrary sequences. The sequence-induced incompatibility between compositionally identical copolymers described by eq.~\ref{eq:chi-final} drives their macroscopic phase separation. The blend free energy can be written as the sum of the translational entropy of chains and their fluctuation-induced pairwise repulsions:
\begin{equation}
    F_{tot} = \frac{\phi_{\alpha}}{N_{\alpha}} \ln \phi_{\alpha} + 
   \frac {\phi_{\beta}} {N_{\beta}} \ln \phi_{\beta} 
    +  \chi_{\alpha \beta}^{RPA} \phi_{\alpha} \phi_{\beta}
\end{equation}
The respective critical point coordinate is given by~\cite{gennes-book, GK-book, RC-book}
\begin{equation}
    \chi_{\alpha \beta}^{RPA} =
    \left. \frac {\left( \sqrt{N_{\alpha}} + \sqrt{N_{\beta}} \right)^2} 
    {2 N_{\alpha} N_{\beta}} \right|_{N_{\alpha} = N_{\beta} = N} = \frac{2}{N}
\end{equation}
Combining the result for length-symmetric blends with eq.~\ref{eq:chi-final} demonstrates that the critical imcompatibility of A and B monomers
\begin{equation}
    \chi_{AB}^{cr} = \frac{4 \pi} {\sqrt{N}} 
    \left[ a^3 \int \left( \frac{ \partial R(q) } { \partial \phi_{\alpha} } \right)^2 q^2 dq \right]^{-1/2} \sim \frac{1}{\sqrt{N}}
\label{eq:scaling}
\end{equation}
scales inversely with the square root of the chain length $N$. Note that the exponent $-1/2$ is different from the $-1$ value for the blend of A and B homopolymers where $\chi_{AB}^{cr} = 2/N$, see eq.~\ref{eq:start}. We remind that the result of eq.~\ref{eq:scaling} is universal and applies to arbitrary copolymers comprising many blocks.

\begin{figure}
\includegraphics[width=3.25in]{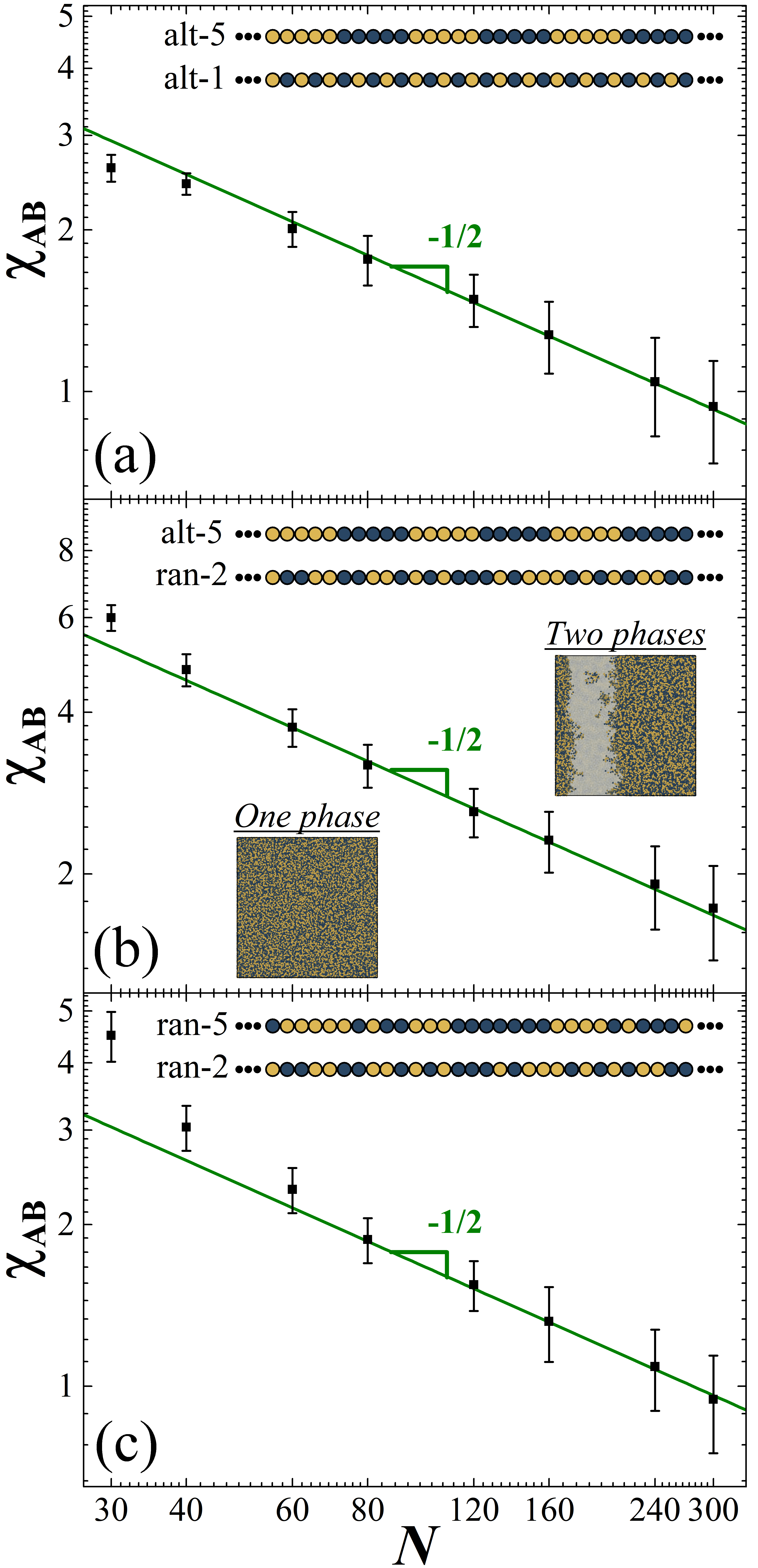}
\caption{
\label{fig:2} 
Simulation dependence of the critical Flory-Huggins parameter $\chi_{AB}^{cr}$ on the length $N$ in a binary blend of compositionally identical AB copolymers with different sequences. The insets in panel (b) illustrate the observed liquid-liquid macrophase separation; only $\beta$ chains are rendered opaque, while $\alpha$ chains are semi-transparent for visual clarity.
} 
\end{figure}

To test our theoretical predictions, we performed dissipative particle dynamics (DPD) simulations~\cite{EW-1995, GW-1997} of copolymer blends. A detailed description of the simulation procedure is provided in the Supporting Information. To demonstrate that the predicted scaling law given by eq.~\ref{eq:scaling} holds universally for chains with both regular and random monomer statistics, we studied three different binary blends of compositionally symmetric AB copolymers: (a) block-alternating copolymers ``alt-5'' with $m = 5$ and ``alt-1'' with $n = 1$; 
(b) a mixture of regular block-alternating copolymers ``alt-5'' with $m = 5$ and ideally random copolymers ``ran-2'' with the average block length of $2$;
(c) statistical (Markov) copolymers with random correlated sequences ``ran-5'' and ideally random copolymers ``ran-2'' with average block lengths of $5$ and $2$, correspondingly.
 For statistical copolymers, only chains with strictly equal 50\%/50\% compositions have been used to avoid phase separation caused by compositional asymmetry and to focus solely on sequence effects.

Figure~\ref{fig:2} shows the dependence of the critical incompatibility $\chi_{AB}^{cr}$ of A and B monomers --- value at which macroscopic phase separation occurs in a 1:1 blend --- on the length $N$ of $\alpha$ and $\beta$ chains. For all three blends, the results confirm the theoretical scaling of $-1/2$.

Some deviations are observed for short chains with $N = 20$ and $30$ because of the end effects. For example, block-alternating copolymers with $m = 5$ consist of only $4$ and $6$ blocks, so the condition $N/m \gg 1$ is not satisfied. This is similar to the dependence of the MST position in block-alternating copolymers on the number of blocks, $N/m$. In the $N/m \to \infty$ limit, the transition occurs at $\chi_{AB} m = 7.55$, whereas tetra- and hexablock copolymers show deviations of 10\% and 6\% from this limiting value~\cite{bates-2004}.

\textbf{Relevance to Intracellular Organization.} The predicted effect of purely sequence-induced demixing in solvent-free blends of $\alpha$ and $\beta$ chains also persists in their solutions at $\phi_{\mathrm{tot}}<1$. In this case, a simple renormalization of the Flory–Huggins interaction parameter in the spirit of the dilution approximation,~\cite{FL-1989} $\chi \to \phi_{\mathrm{tot}}\chi$, is sufficient.~\cite{rumyantsev-2024} This is because the $\alpha$–$\beta$ phase separation is driven by fluctuations of the compositional field $\psi (\vec{r})$, whereas fluctuations of the orthogonal order parameter, the total polymer density $\phi_{tot} (\vec{r})$, do not play an essential role in the demixing mechanism.~\cite{rumyantsev-2024} This renders our results directly relevant for intracellular liquid–liquid phase separation.

Studies of intrinsically disordered proteins have clearly shown that primary sequence plays a crucial role in determining their propensity for phase separation and membraneless organelles formation~\cite{nott-2015, mittal-2020, pappu-2020, webb2024}. In particular, the patterning of charged residues interacting electrostatically~\cite{nott-2015, PRL-2017, fredrickson-2019} and non-ionic aromatic residues acting as hydrophobic stickers~\cite{pappu-2020, statt-2020} has emerged as the dominant factor. The effect of sequence on multiphase coexistence within biocondensates~\cite{brangwynne-2021} remains an open question, and our results help to answer it from a polymer physics perspective~\cite{pappu-2015}.

The present theory assumes that interactions between A and B monomers are short-range. It can be applied to not only non-ionic AB hydrophilic-hydrophobic copolymers but also polyampholytes of $A^{+}$ and $B^{-}$ monomers, provided salt concentration is sufficiently high --- in this case, Coulomb interactions between the charges become effectively short-range pairwise~\cite{RJ-2023}. Interestingly, in view of the physiological salt concentration of $\approx 0.15$M within the cell~\cite{cell-salinity}, the model with short-range interactions may be better suited for multiphase condensates than one with long-range (bare) Coulomb forces~\cite{yang-2023}. 

In this Letter, we develop the RPA theory of macroscopic phase separation between AB copolymers that are identical in overall composition but differ only in primary sequence. Their incompatibility arises from distinct positional correlations between long blocks of $\alpha$ chains versus short blocks of $\beta$ chains. Strikingly, this second-order fluctuation-induced effect yields the new scaling slope of $-1/2$ for the critical FH parameter $\chi_{AB}^{cr}$ against chain length $N$, $\chi_{AB}^{cr} \sim 1 / \sqrt{N}$, in contrast to the standard $-1$ for chains of different compositions. The universality of this law, confirmed by simulations for periodic and statistical copolymers, establishes a minimal model for intracellular multiphase coexistence driven solely by sequence variation.

\subsection*{Supporting Information}

1. Inverse matrix of the Fourier transforms of bare structure factors,  $\textbf{G}_{0}^{-1} (q)$.
2. Exact result and accuracy of analytical approximation for blends of block-alternating copolymers. 
3. Details of dissipative particle dynamics (DPD) simulations.

\subsection*{Acknowledgements}

This work used Jetstream2 CPU at Indiana University through allocation PHY240314 from the Advanced Cyberinfrastructure Coordination Ecosystem: Services \& Support (ACCESS) program~\cite{boernerACCESSAdvancingInnovation2023}, which is supported by U.S. National Science Foundation grants \#2138259, \#2138286, \#2138307, \#2137603, and \#2138296. This work also used the computing resources provided by North Carolina State University High Performance Computing Services Core Facility.

\bibliography{references}

@PREAMBLE{
 "\providecommand{\noopsort}[1]{}" 
 # "\providecommand{\singleletter}[1]{#1}%" 
}

@ARTICLE{huggins-1942,
   author       = "M. L. Huggins",
   title        = "Theory of Solutions of High Polymers",
   year         = "1942",
   journal      = "J. Am. Chem. Soc.",
   volume       = "64",
   pages        = "1712-1719",
}

@ARTICLE{flory-1942,
   author       = "P. J. Flory",
   title        = "Thermodynamics of High Polymer Solutions",
   year         = "1942",
   journal      = "J. Chem. Phys.",
   volume       = "10",
   pages        = "51-61",
}

@ARTICLE{scott-1952,
   author       = "R. L. Scott",
   title        = "Thermodynamics of High Polymer Solutions. VI. The Compatibility of Copolymers",
   year         = "1955",
   journal      = "J. Polym. Sci.",
   volume       = "9",
   pages        = "423-432",
}

@ARTICLE{ten-brinke-1983,
   author       = "G. ten Brinke and F. E. Karasz and W. J. MacKnight",
   title        = "Phase Behavior in Copolymer Blends: Poly(2,6-Dimethyl-1,4-Phenylene Oxide) and Halogen-Substituted Styrene Copolymers",
   year         = "1983",
   journal      = "Macromolecules ",
   volume       = "16",
   pages        = "1827–1832",
}

@BOOK{flory-book,
   author       = {P. J. Flory},
   year         = 1953,
   title        = {Principles of Polymer Chemistry},
   publisher    = {Cornell University Press}
}

@BOOK{gennes-book,
   author       = {P-G. de Gennes},
   year         = 1979,
   title        = {Scaling Concepts in Polymer Physics},
   publisher    = {Cornell University Press}
}

@BOOK{GK-book,
   author       = {A. Yu. Grosberg and A. R. Khokhlov},
   year         = 1994,
   title        = {Statistical Physics of Macromolecules},
   publisher    = {AIP Press}
}

@BOOK{RC-book,
   author       = {M. Rubinstein and R. H. Colby},
   year         = 2003,
   title        = {Polymer Physics},
   publisher    = {Oxford University Press}
}

@BOOK{F-book,
   author       = {G. H. Fredrickson},
   year         = 2006,
   title        = {The Equilibrium Theory of Inhomogeneous Polymers},
   publisher    = {Oxford University Press}
}

@BOOK{Landau-book,
   author       = {L. D. Landau and E. M. Lifshitz},
   year         = 1980,
   title        = {Statistical Physics, Part 1},
   publisher    = {Butterworth–Heinemann, Oxford}
}

@ARTICLE{leibler-1980,
   author       = "L. Leibler",
   year         = "1980",
   title        = {Theory of Microphase Separation in Block Copolymers},
   journal      = "Macromolecules",
   volume       = "13",
   pages        = "1602-1617",
}

@ARTICLE{erukh-1982,
   author       = "I. Ya. Erukhimovich",
   year         = "1982",
   title        = {Fluctuations and the Formation of Domain Structure in Heteropolymers},
   journal      = "Polym. Sci. U.S.S.R.",
   volume       = "24",
   pages        = "2223-2232",
}

@ARTICLE{morse-2010,
   author       = "J. Qin and F. S. Bates and D. C. Morse",
   year         = "2010",
   title        = {Phase Behavior of Nonfrustrated ABC Triblock Copolymers: Weak and Intermediate Segregation},
   journal      = "Macromolecules",
   volume       = "43",
   pages        = "5128-5136",
}

@ARTICLE{OK-1986,
   author       = "T. Ohta and K. Kawasaki",
   year         = "1986",
   title        = {Theory of Block Copolymer Solutions: Nonselective Good Solvents},
   journal      = "Macromolecules",
   volume       = "19",
   pages        = "2621-2632",
}

@ARTICLE{FL-1989,
   author       = "G. Fredrickson and L. Leibler",
   year         = "1989",
   title        = {Theory of Block Copolymer Solutions: Nonselective Good Solvents},
   journal      = "Macromolecules",
   volume       = "22",
   pages        = "1238-1250",
}

@ARTICLE{edwards-1966,
   author       = "S. F. Edwards ",
   year         = "1966",
   title        = {The theory of polymer solutions at intermediate concentration},
   journal      = "Proc. Phys. Soc.",
   volume       = "88",
   pages        = "265-280"
}

@ARTICLE{olvera-1988,
   author       = "M. Olvera de la Cruz and S. F. Edwards and I. C. Sanchez",
   year         = "1988",
   title        = {Concentration fluctuations in polymer blend thermodynamics},
   journal      = "J. Chem. Phys.",
   volume       = "89",
   pages        = "1704-1708",
}

@ARTICLE{erukh-1988,
   author       = "V. Yu. Borue and I. Ya. Erukhimovich",
   year         = "1988",
   title        = { A Statistical Theory of Weakly Charged Polyelectrolytes: Fluctuations, Equation of State, and Microphase Separation},
   journal      = "Macromolecules",
   volume       = "21",
   pages        = "3240-3249",
}

@ARTICLE{BF-1994,
   author       = "Bates, F. S. and Fredrickson, G. H.",
   year         = "1994",
   title        = {Conformational Asymmetry and Polymer-Polymer Thermodynamics},
   journal      = "Macromolecules ",
   volume       = "27",
   pages        = "1065–1067",
}

@ARTICLE{BLF-1994,
   author       = "Fredrickson, G. H. and Liu, A. J. and Bates, F. S. ",
   year         = "1994",
   title        = {Entropic Corrections to the Flory-Huggins Theory of Polymer Blends: Architectural and Conformational Effects},
   journal      = "Macromolecules ",
   volume       = "27",
   pages        = "2503–2511",
}

@ARTICLE{potemkin-2007,
   author       = "N. N. Oskolkov and I. I. Potemkin",
   year         = "2007",
   title        = {Complexation in Asymmetric Solutions of Oppositely Charged Polyelectrolytes: Phase Diagram},
   journal      = "Macromolecules",
   volume       = "40",
   pages        = "8423-8429",
}

@ARTICLE{joanny-2002,
   author       = "M. Castelnovo and J.-F. Joanny",
   year         = "2001",
   title        = {Complexation between oppositely charged polyelectrolytes: Beyond the Random Phase Approximation},
   journal      = "Eur. Phys. J. E",
   volume       = "6",
   pages        = "377-386",
}

@ARTICLE{RJTdP-2022,
   author       = "A. M. Rumyantsev and M. V. Tirrell and  A. Johner and J. J. de Pablo",
   year         = "2022",
   title        = {Unifying Weak and Strong Charge Correlations within the Random Phase Approximation: Polyampholytes of Various Sequences},
   journal      = "Macromolecules",
   volume       = "55",
   pages        = "6260-6274",
}

@ARTICLE{RJ-2023,
   author       = "A. M. Rumyantsev and A. Johner",
   year         = "2023",
   title        = {Salt-added solutions of Markov polyampholytes: Diagram of states, antipolyelectrolyte effect, and self-coacervate dynamics},
   journal      = "Macromolecules",
   volume       = "56",
   pages        = "5201-5216",
}

@ARTICLE{rumyantsev-2024,
   author       = "A. M. Rumyantsev",
   year         = "2024",
   title        = {Why Sequence Blockiness Sharpens Coil-Globule Transition in Heteropolymers},
   journal      = "Macromolecules",
   volume       = "57",
   pages        = "10454-10462",
}

@ARTICLE{wang-2002,
   author       = "Z.-G. Wang",
   year         = "2002",
   title        = {Concentration fluctuation in binary polymer blends: $\chi$ parameter, spinodal and Ginzburg criterion},
   journal      = "J. Chem. Phys.",
   volume       = "117",
   pages        = "481-500",
}

@ARTICLE{morse-2007,
   author       = "P. Grzywacz and J. Qin and D. C. Morse",
   year         = "2007",
   title        = {Renormalization of the one-loop theory of fluctuations in polymer blends and diblock copolymer melts},
   journal      = "Phys. Rev. E",
   volume       = "76",
   pages        = "061802",
}

@ARTICLE{rauscher-2023,
   author       = "P. M. Rauscher",
   year         = "2023",
   title        = {Renormalized one-loop theory of correlations in disperse polymer blends},
   journal      = "J. Chem. Phys.",
   volume       = "159",
   pages        = "244906",
}

@ARTICLE{bates-2004,
   author       = "Wu, L. and Cochran, E. W. and Lodge, T. P. and Bates, F. S.",
   year         = "2004",
   title        = {Consequences of block number on the order-disorder transition and viscoelastic properties of linear $(\text{AB})_{n}$  multiblock copolymers},
   journal      = "Macromolecules",
   volume       = "37",
   pages        = "3360--3368",
}

@ARTICLE{cell-salinity,
   author       = "Liu, B. and Poolman, B. and Boersma, A. J.",
   year         = "2017",
   title        = {Ionic Strength Sensing in Living Cells},
   journal      = "ACS Chem. Biol.",
   volume       = "12",
   pages        = "2510--2514",
}

@ARTICLE{yang-2023,
   author       = "X. Chen and E.-Q. Chen and S. Yang",
   year         = "2023",
   title        = {Multiphase Coacervation of Polyelectrolytes Driven by Asymmetry
of Charged Sequence},
   journal      = "Macromolecules",
   volume       = "56",
   pages        = "3-14",
}

@ARTICLE{nott-2015,
   author       = "Nott, T. J. and Petsalaki, E. and Farber, P. and Jervis, D. and Fussner, E. and Plochowietz, A. and Craggs, T. D. and Bazett-Jones, D. P. and Pawson, T. and Forman-Kay, J. D. and Baldwin, A. J.",
   year         = "2015",
   title        = {Phase Transition of a Disordered Nuage Protein Generates Environmentally Responsive Membraneless Organelles},
   journal      = "Mol. Cell.",
   volume       = "57",
   pages        = "936-947",
}

@ARTICLE{PRL-2017,
   author       = "Y.-H. Lin and J. D. Forman-Kay and H. S. Chan",
   year         = "2016",
   title        = {Sequence-Specific Polyampholyte Phase Separation in Membraneless Organelles},
   journal      = "Phys. Rev. Lett.",
   volume       = "117",
   pages        = "178101",
}

@ARTICLE{fredrickson-2019,
   author       = "McCarty, J. and Delaney, K. T. and Danielsen, S. P. O. and Fredrickson, G. H. and Shea, J.-E.",
   year         = "2019",
   title        = {Complete Phase Diagram for Liquid–Liquid Phase Separation of Intrinsically Disordered Proteins},
   journal      = "J. Phys. Chem. Lett.",
   volume       = "20",
   pages        = "1644–1652",
}

@ARTICLE{statt-2020,
   author       = "Statt, A. and Casademunt, H. and Brangwynne, C. P. and Panagiotopoulos, A. Z.",
   year         = "2020",
   title        = {Model for Disordered Proteins with Strongly Sequence-Dependent Liquid Phase Behavior},
   journal      = "J. Chem. Phys.",
   volume       = "152",
   pages        = "075101",
}

@ARTICLE{pappu-2020,
   author       = "Martin, E. W. and Holehouse, A. S. and Peran, I. and Farag, M. and Incicco, J. J. and Bremer, A. and Grace, C. R. and Soranno, A. and Pappu, R. V. and Mittag, T.",
   year         = "2020",
   title        = {Valence and Patterning of Aromatic Residues Determine the Phase Behavior of Prion-like Domains},
   journal      = "Science",
   volume       = "367",
   pages        = "694-699",
}

@ARTICLE{mittal-2020,
   author       = "Schuster, B. S. and Dignon, G. L. and Tang, W. S. and Kelley, F. M. and Ranganath, A. K. and Jahnke, C. N. and Simpkins, A. G. and Regy, R. M. and Hammer, D. A. and Good, M. C. and Mittal, J.",
   year         = "2020",
   title        = {Identifying Sequence Perturbations to an Intrinsically Disordered Protein That Determine Its Phase-Separation Behavior},
   journal      = "Proc. Natl. Acad. Sci. U.S.A.",
   volume       = "117",
   pages        = "11421-11431",
}

@ARTICLE{pappu-2015,
   author       = "Brangwynne, C. P. and Tompa, P. and Pappu, R. V.",
   year         = "2015",
   title        = {Polymer Physics of Intracellular Phase Transitions},
   journal      = "Nature Phys.",
   volume       = "11",
   pages        = "899-904",
}

@ARTICLE{brangwynne-2021,
   author       = "Lafontaine, D. L. J. and Riback, J. A. and Bascetin, R. and Brangwynne, C. P.",
   year         = "2021",
   title        = {The Nucleolus as a Multiphase Liquid Condensate. },
   journal      = "Nature Rev. Mol. Cell. Biol.",
   volume       = "22",
   pages        = "165–182",
}

@ARTICLE{webb2024,
  author       = {Yaxin An and Michael A. Webb and William M. Jacobs},
  title        = {Active learning of the thermodynamics–dynamics trade-off in protein condensates},
  journal      = {Science Advances},
  year         = {2024},
  volume       = {10},
  number       = {1},
  pages        = {eadj2448},
}

@ARTICLE{EW-1995,
   author       = "Español, P. and Warren, P. ",
   year         = "1995",
   title        = {Statistical Mechanics of Dissipative Particle Dynamics},
   journal      = "Europhys. Lett.",
   volume       = "30",
   pages        = "191–196",
}

@ARTICLE{GW-1997,
   author       = "Groot, R. D. and Warren, P. B.",
   year         = "1997",
   title        = {Dissipative Particle Dynamics: Bridging the Gap between Atomistic and Mesoscopic Simulation},
   journal      = "J. Chem. Phys.",
   volume       = "107",
   pages        = "4423–4435",
}

@article{Levin2002,
author = {Levin, Yan},
issn = {0034-4885},
journal = {Reports on Progress in Physics},
month = {nov},
number = {11},
pages = {1577--1632},
title = {{Electrostatic correlations: from plasma to biology}},
url = {http://stacks.iop.org/0034-4885/65/i=11/a=201?key=crossref.d7092a21b21172a84f5b57a377177846},
volume = {65},
year = {2002}
}

@article{boernerACCESSAdvancingInnovation2023,
  title = {{{ACCESS}}: {{Advancing Innovation}}: {{NSF}}'s {{Advanced Cyberinfrastructure Coordination Ecosystem}}: {{Services}} \& {{Support}}},
  shorttitle = {{{ACCESS}}},
  booktitle = {Practice and {{Experience}} in {{Advanced Research Computing}} 2023: {{Computing}} for the {{Common Good}}},
  author = {Boerner, Timothy J. and Deems, Stephen and Furlani, Thomas R. and Knuth, Shelley L. and Towns, John},
  year = 2023,
  month = sep,
  series = {{{PEARC}} '23},
  pages = {173--176},
  publisher = {Association for Computing Machinery},
  address = {New York, NY, USA},
  urldate = {2026-01-19},
  isbn = {978-1-4503-9985-2},
}

\end{document}